# Synchronization of chiral vortex nano-oscillators

Zhiyang Zeng[1,2,3], Zhaochu Luo[1,2], Laura J. Heyderman[1,2], Joo-Von Kim[4], Aleš Hrabec[1,2,5,*]

[1]Laboratory for Mesoscopic Systems, Department of Materials, ETH Zurich, 8093 Zurich, Switzerland

[2]Laboratory for Multiscale Materials Experiments, Paul Scherrer Institute, 5232 Villigen PSI, Switzerland

[3]Yuanpei College, Peking University, Beijing 100871, People's Republic of China

[4]Centre de Nanosciences et de Nanotechnologies, CNRS, Université Paris-Saclay, 91120 Palaiseau, France

[5]Laboratory for Magnetism and Interface Physics, Department of Materials, ETH Zurich, 8093 Zurich, Switzerland

*ales.hrabec@psi.ch

## Abstract

The development of spintronic oscillators is driven by their potential applications in radio frequency telecommunication and neuromorphic computing. In this work, we propose a spintronic oscillator based on the chiral coupling in thin magnetic films with patterned anisotropy. With an in-plane magnetized disk imprinted on an out-of-plane magnetized slab, the oscillator takes a polar vortex-like magnetic structure in the disk stabilized by a strong Dzyaloshinskii-Moriya interaction. By means of micromagnetic simulations, we investigate its oscillatory properties under applied spin current and, by placing an ensemble of oscillators in the near vicinity, we demonstrate their synchronization with different resonant frequencies. Finally, we show their potential application in neuromorphic computing using a network with six oscillators.

## Main text

Spin-torque nano-oscillators are based on the self-oscillatory motion of magnetization whose operational frequencies, determined by competing micromagnetic torques, span a broad range up to several tens of GHz[1]. The magnetic oscillators can involve various forms, including magnetic textures defined by lithographic patterning[2–5], naturally occurring spin configurations in magnetic films[6–11] or in a uniform magnetic state that is brought out of the equilibrium[12–14]. Magnetization oscillations can be driven by both magnetic field and electric current, while the latter offers more attractive ways to control them locally with better scalability. Crucially, sustained auto-oscillations can be achieved when the intrinsic damping is compensated over an oscillation period by the resulting current-induced torques[15,16]. Importantly for technological applications, such oscillations can be translated to a high frequency electric signal using the giant or tunnel magnetoresistance effect. Such oscillators are miniaturized high-frequency counterparts of widely used radio-frequency (RF) oscillators.

A general property of oscillators is their capacity to synchronize with one another via mutual interactions[17]. For magnetic oscillators, the coupling can be mediated via overlapping magnetic structures[18,19], the dipolar interaction[20–22], spin waves[23] or via emitted electric signals[24]. Apart from the fundamental interest for nonlinear dynamics, this effect broadens the oscillators' functionalities. Firstly, an array of synchronized oscillators provides a means to enhance the weak output power of a single oscillator. Secondly, the rich nonlinear dynamics of spintronic oscillators can be applied to neuromorphic computing, whose aim is to provide low-power solutions to cognitive tasks such as pattern recognition with noisy data[25]. Exploiting the synchronization property of spintronic nano-oscillators, the bio-inspired computation scheme combines both computation and memory at the local level, which can potentially decrease the vast gap in power consumption between artificial computers and human brains[26]. Up to now, various architectures have been developed to demonstrate the feasibility and superiority of spintronic oscillator-based neuromorphic computing in the applications of pattern recognition and classification[27–30].

Here, we propose a design of spintronic nano-oscillator based on in-plane (IP) magnetized regions in a perpendicular magnetic anisotropy (PMA) material in combination with an interfacial Dzyaloshinskii-Moriya interaction (DMI). We present a systematic study of the DMI-based spintronic oscillator using micromagnetic simulations. By imprinting the magnetic configuration of an IP magnetized disk in an out-of-plane (OOP) magnetized slab containing sizeable DMI, a chiral vortex texture[31] can be established as a ground state in the patterned region, which is confined by a 90° domain wall as the magnetization transitions to the perpendicularly magnetized state outside this region. We show that such an oscillator can be driven and manipulated by an injected spin current. We also investigate the mutual synchronization of such oscillators via spin wave and dipolar interactions, and demonstrate the potential for neuromorphic computing with a neural network of six oscillators.

The design of our oscillator is inspired by previous experimental studies on chirally coupled nanomagnets[32]. The discovery of chiral coupling between IP and OOP magnetized regions in a thin film by DMI led to the realization of artificial spin ices[33], lateral synthetic antiferromagnets, field-free memory elements[32] and current-driven domain wall injectors[34] as well as domain wall logic circuits[35–37]. In these systems with both an asymmetric structure environment and non-uniform magnetic anisotropy, antisymmetric exchange coupling, also known as DMI, tends to stabilize chirally coupled magnetization textures. The fabrication of chirally coupled nanomagnets are normally based on selective oxidation of the top Al film in Pt/Co/Al textures and can be extended to other materials interesting for high frequency applications by different anisotropy patterning methods[36,38–41]. The additional uniaxial anisotropy contribution arising at the Co/AlOx interface[42] enables the patterning of a specific region with OOP magnetization. The DMI arising at the Pt/Co interface chirally couples the magnetization at the interface between IP and OOP region. Given the DMI sign in Pt/Co/Al films[32,43], the magnetic films have a preferred left-handed chirality. Based on this principle, we create a circular IP magnetization region in the film, surrounded by an OOP region, which serves as the basic structure of the oscillator [Fig. 1(a, b)]. The magnetization in the IP region assumes a polar vortex state with a chirality fixed by the DMI[44]. The vortex is confined in the IP magnetized region and is bounded by 90° chiral domain walls as the magnetization transitions toward a perpendicular orientation outside the region.

The micromagnetic simulations are performed with MuMax3[45] with the cell sizes of 2 nm x 2 nm x 1 nm. Magnetic parameters used in the simulations are based on the prototypical Pt/Co/AlOx trilayer film, where we assume a saturation magnetization of $M_s = 1$ MA/m, an exchange constant of $A = 16$ pJ/m, a uniaxial anisotropy constant of $K_0 = 0.703$ MJ/m$^3$ (resulting in an effective anisotropy field of 150 mT) in the OOP region, and $\alpha = 0.01$ for the Gilbert damping parameter. The interfacial DMI strength is set to be $D = -1.5$ mJ/m$^2$ (unless otherwise specified), which favors left-handed Néel states. Prior to each run of the simulation, the initial state is reached by relaxing the system from an OOP magnetization in the OOP region and an opposite magnetization in the IP region. Spin current polarized along the *z*-axis is injected after the system reaches its initial state. The oscillation frequencies of the studied cases are extracted by performing a fast Fourier transformation (FFT) from the time-dependent *x*-component of the averaged magnetization in the IP region over an interval of 150 ns. The frequency spectrum of two oscillators are superimposed to study synchronization phenomena. The combined spectra of the two oscillators exhibit a single peak when mutual phase-locking occurs, otherwise two separate peaks can be discerned. The simulations were performed at zero temperature and thermal magnetization fluctuations were not taken into account.

Due to the symmetry of the system, there are two equivalent ground states of the oscillator, depending on the magnetization direction of the OOP region. When the initial magnetization in OOP region is set in the +*z* direction [Fig. 1(a)], the relatively strong OOP anisotropy stabilizes the uniform magnetization in the outside region while the magnetic texture in the IP magnetized disk is radial with a core pointing in the -*z* direction, which is stabilized primarily by the interfacial DMI and exchange interaction. When the initializing OOP magnetic field is set along the opposite direction (-*z*), both the IP magnetization orientation within the disk and the direction of the core are reversed [Fig. 1(b)], satisfying the unique winding sense set by DMI.

We first study the equilibrium position of the vortex core by calculating the relative potential energy of the system with different DMI strengths and the core at different distances from the center [Fig. 1(c)]. For a relatively large DMI strength ($\geq$ 0.75 mJ/m$^2$), the minimum in the potential energy corresponds to the case where the core is at the center of the patterned region while at lower DM strengths ($\leq$ 0.5 mJ/m$^2$), the minimum potential energy occurs at a finite distance of the vortex core from the center of the patterned disk. This supports the observation that the equilibrium position of the core moves to the center when increasing the DMI, resulting in a radial vortex structure[46]. In what follows, we will focus on the cases with a relatively large DMI strength ($\geq$ 0.75 mJ/m$^2$) as they result in a larger frequency tunability with applied currents.

In order to quantify the form of the confining potential, we applied a lateral magnetic field to the equilibrium state, which serves to displace the core away from its equilibrium position. Once the magnetic field is removed, the core relaxes back towards its initial equilibrium position through damped gyrotropic motion. The motion of the core can be described by Thiele equation[21,47,48]

$$\vec{G} \times \dot{\vec{R}} + k(\vec{R})\vec{R} + \alpha\eta G\vec{R} - F_{STT}(\vec{R}) = 0, \quad (1)$$

where $\vec{G} = -2\pi\hat{z}PM_s d/\gamma$ is the gyrovector, $P$ is the core polarity, $M_s$ is the saturation magnetization, $d$ is the film thickness, $\gamma$ is the gyromagnetic ratio, $\vec{R}$ is the core position vector, $\alpha$ is the Gilbert damping constant, $\eta$ is a phenomenological damping coefficient for the vortex, and $F_{STT}(\vec{R})$ is the spin-transfer force. The restoring force, $k$, is based on a confining potential, $U(R)$, defined by the IP magnetized region, $k(R) = -\partial_R U(R)$.

The intrinsic frequency of the oscillation is determined by the curvature of the confining potential at the equilibrium core position in the absence of spin torques. This curvature possesses a strong dependence on the diameter of the oscillator and the interfacial DMI strength. To demonstrate the impact of the geometry and DMI, we have simulated oscillators with diameters ranging from 100 to 300 nm and with DMI strength from 0.8 to 2.5 mJ/m$^2$. As can be seen in Figs. 1(d) and 1(e), the resonant frequency of the oscillator increases with a reduced size and with increased DMI strength.

Due to the dissipation term in Equation (1), the amplitude of the oscillation decays over time during the oscillation and the core gradually returns to the center. Therefore, a continuous energy flow is required to counteract the dissipation such that continuous gyration of the vortex core can be sustained, which is required for applications. This can be achieved by injecting a spin current polarized in the *z* direction to the oscillator with an OOP layer on the top[49] [Fig. 1(f)]. It is shown in Fig. 1(g) that the oscillations can be successfully sustained with the injected current[50]. For applied currents below the threshold for self-sustained core gyration, only damped oscillations in the core position are observed. We can determine the resonance frequency of the system from these oscillations. When the current threshold is reached, self-sustained gyrations of the vortex core can be observed, which take place at a finite distance $R_0$ from the center of the patterned region. As the current is increased, the radius of the steady state orbit increases as the larger dissipation rate is required to compensate the increase in spin torques. This entails higher oscillation frequencies [Fig. 1(h)] above the intrinsic resonance frequency, which corresponds to a current-induced blue-shift as seen in other vortex-based oscillators. This variation can be detected as a change in the overall IP magnetization through the giant- or tunnel magnetoresistance effect.

Due to the mutual interactions, two spintronic oscillators with different intrinsic frequencies can phase-lock to each other if their oscillation frequencies are within a certain locking range. The same phenomenon is observed in our DMI-based oscillators, when two identical oscillators are placed at a given distance and they are driven with spin currents of different intensities [Fig. 2(a)]. Here, the driving current density of one oscillator is fixed at 0.8×10$^9$ A/m$^2$ while that of the other oscillator is swept from 0.1 to 1.35×10$^9$ A/m$^2$. It can be seen in Fig. 2(a) that the two oscillators spaced 60 nm apart can synchronize to each other when the current density of the second oscillator ranges from 0.2 × 10$^9$ to 1.35 × 10$^9$ A/m$^2$. The synchronization range becomes narrower when the spacing is increased from 60 nm to 100 nm [Fig. 2(b)]. This decrease of the synchronization range due to decay of the coupling strength with the increased distance is shown systematically in Fig. 2(c).

In order to investigate whether the coupling between the oscillators is mainly mediated by propagating spin waves or by dipolar coupling, we introduced a trench between two oscillators to prevent possible propagating spin waves[51] [Fig. 3(a)]. The trench is set as a vacuum in the simulation. It turns out that the synchronization is maintained with an approximately same synchronization range after introducing the trench [Fig. 3(b)], suggesting that the dipolar interaction alone can promote mutual synchronization. To better study the effect of spin waves in the system, we measured the emitted spin waves of a single oscillator at different distances. It is shown in Fig. 4(a) that the amplitude of the emitted spin wave decays with the distance from the edge of the oscillator, as expected from Gilbert damping and propagation losses. Moreover, the decay rate of the spin wave is affected by applying a magnetic field in *z* direction, which shifts the overall spin wave band toward higher frequencies with increasing fields. The suppression of spin wave propagation by the magnetic field also influences the synchronization between two oscillators, which results in a reduced synchronization region in Fig. 4(b) compared to the case without external field [Fig. 2(a)]. The tunability of coupling strength by an applied magnetic field ensures a broader applicability of our oscillators in real devices.

To further demonstrate the application of our DMI-based oscillators in neuromorphic computing, we created a model neural network with six oscillators based on the scheme in Ref. 27 [Fig. 5(a)]. Among them, two oscillators serve as inputs (A and B), whose injected current can be precisely tuned while the injected currents of other four neuron oscillators (1, 2, 3 and 4) are fixed. With different inputs, the network will naturally converge to different synchronization states, which can be taken as output for recognition or classification. For example, when input currents for A and B are set to be 0.94 $\times 10^9$ A/m$^2$ and 0.89 $\times$ 10$^9$ A/m$^2$ respectively, five oscillators (1, 2, 3, A and B) in the network will synchronize together [Fig. 5(b)] and the synchronization pattern is labeled as (123AB). A total of 11 different synchronization patterns can be achieved when sweeping the input currents in the given ranges [Fig. 5(c)], which means the network is capable of distinguishing 11 different categories of inputs.

In summary, we proposed a design of spintronic oscillators based on DMI, which has shown great potential in the applications of neuromorphic computing and tunable spin wave emitters[52,53]. Through a systematic study, we revealed the basic dynamic properties of the oscillator and achieved precise manipulation of the oscillator with spin polarized current and magnetic field. Compared with conventional vortex-based spintronic oscillators, the oscillator we proposed has the advantage of miniaturization and even benefits from downscaling, which leads to an increase in the intrinsic frequencies. Moreover, we studied the synchronization between the oscillators and verified that the coupling mechanism involves both dipolar and spin wave interactions. Based on this synchronization, we built up a six-oscillator system, and demonstrated the feasibility of our oscillators in neuromorphic computing.

**Acknowledgments**

Z.Z. thanks Mr. Wenshuo Yue for the fruitful discussion. A.H. was funded by the European Union's Horizon 2020 research and innovation program under Marie Sklodowska-Curie grant agreement number 794207 (ASIQS). Z.L. and L.J.H. acknowledge funding from the European Union's Horizon 2020 FET-Open program under grant agreement No 861618 (SpinEngine). The data that support this study are available via the Zenodo repository[54].

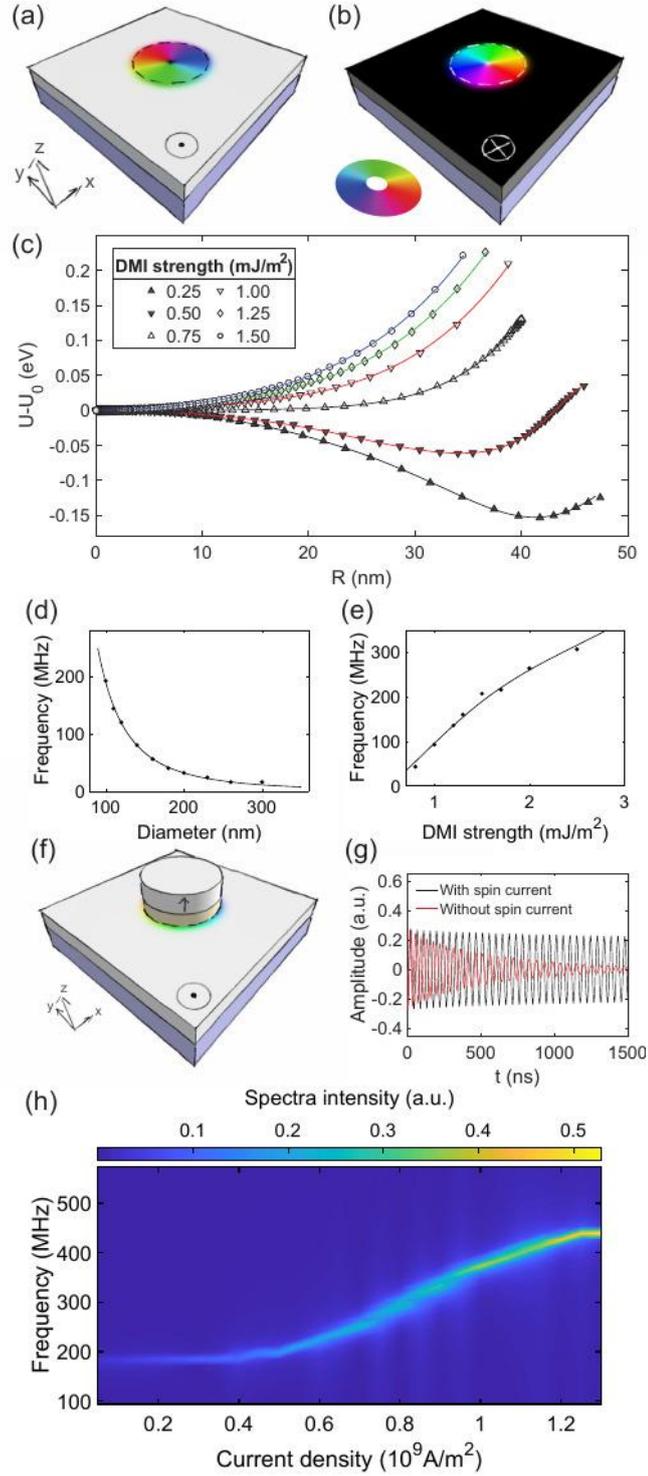

Fig. 1. (a, b) Equilibrium magnetic texture in the DMI-based nano-oscillator with 100 nm diameter. The oscillator has two equivalent states, depending on the magnetization direction in the OOP region. Magnetization along the +z and –z direction are indicated in white and black, while the IP magnetization direction is given by the color wheel. (c) The dependence of potential energy of the system on DMI strength and the position of the core. $R$ is the distance of the core from the center. (d) Dependence of intrinsic oscillation frequency on the oscillator diameter. The DMI strength is set to 1.5 mJ/m². (e) Dependence of intrinsic oscillation

frequency on the DMI strength. The diameter of the patterned region is fixed to 100 nm. (f) Schematic for current-driven auto-oscillations with an OOP spin polarization layer placed on top of the IP region. (g) Oscillation amplitudes of $m_x$ component inside the IP region with (black) and without (red) spin current. (h) Dependence of oscillation frequency on spin current density.

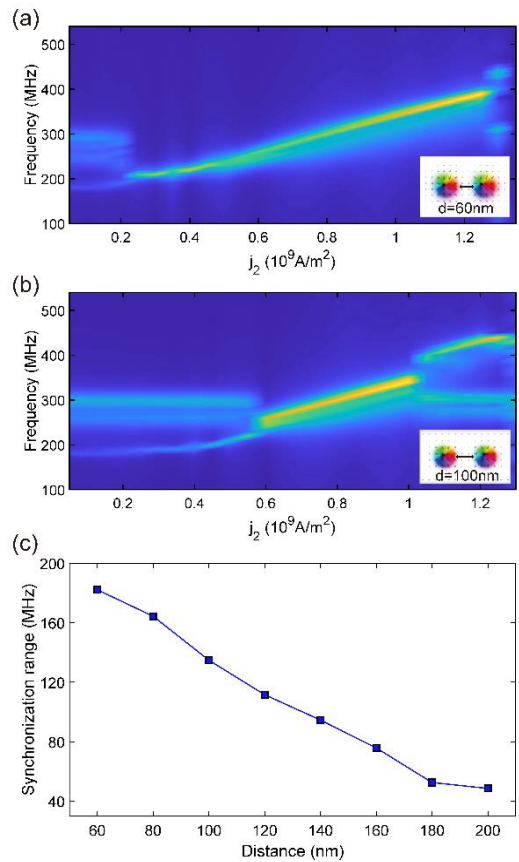

Fig. 2. (a, b) Synchronization pattern of two oscillators separated by (a) 60 nm and (b) 100 nm distance. The current density in the left oscillator is fixed ($j_1 = 0.8 \times 10^9$ A/m²) while the current density of the oscillator on the right $j_2$ is swept. (c) The dependence of synchronization range on the distance between two oscillators.

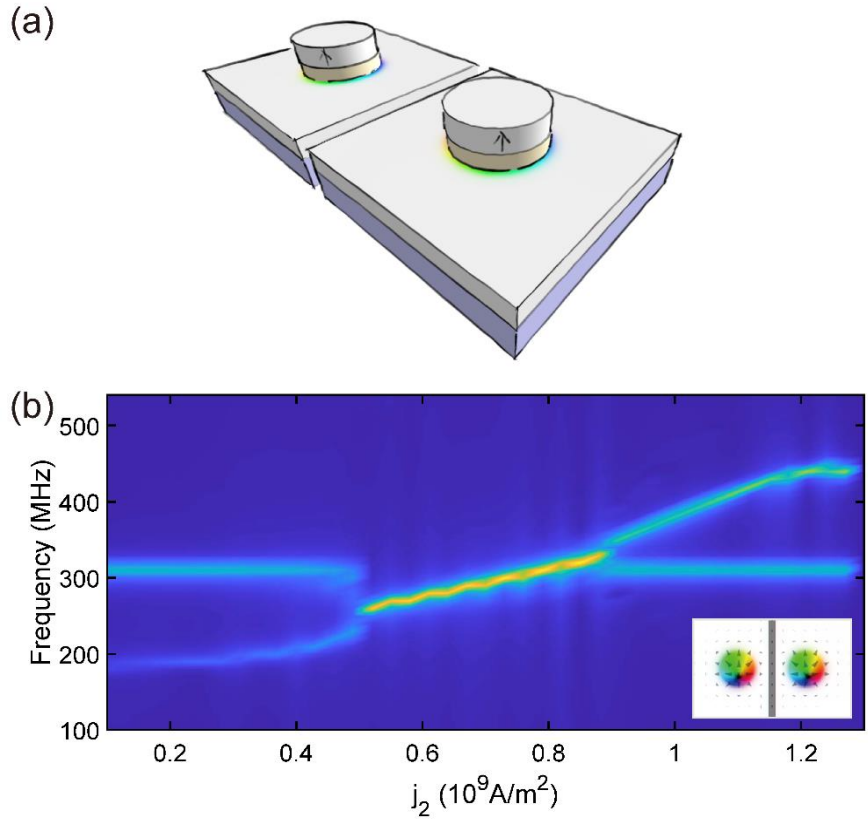

Fig. 3. (a) Schematic of two oscillators separated by a trench. (b) Synchronization pattern of two oscillators separated by 100 nm distance with a 10 nm-wide trench between them. The current density in the left oscillator is fixed ($j_1 = 0.8 \times 10^9$ A/m2) while the current density of the oscillator on the right is varied.

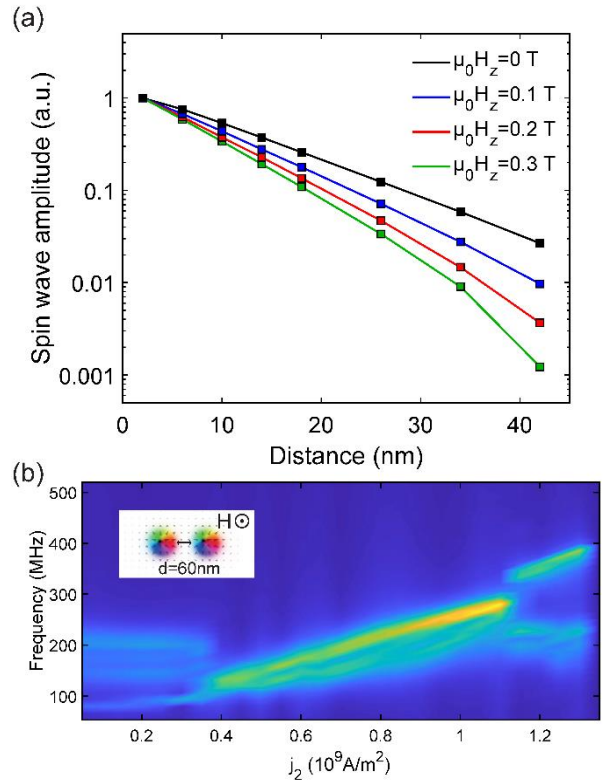

Fig. 4. (a) Normalized spin wave amplitude. The spin wave decays with distance and the decay rate is enhanced with the applied magnetic field. (b) Synchronization pattern of two oscillators separated by 60 nm distance in a 0.3 T applied magnetic field.

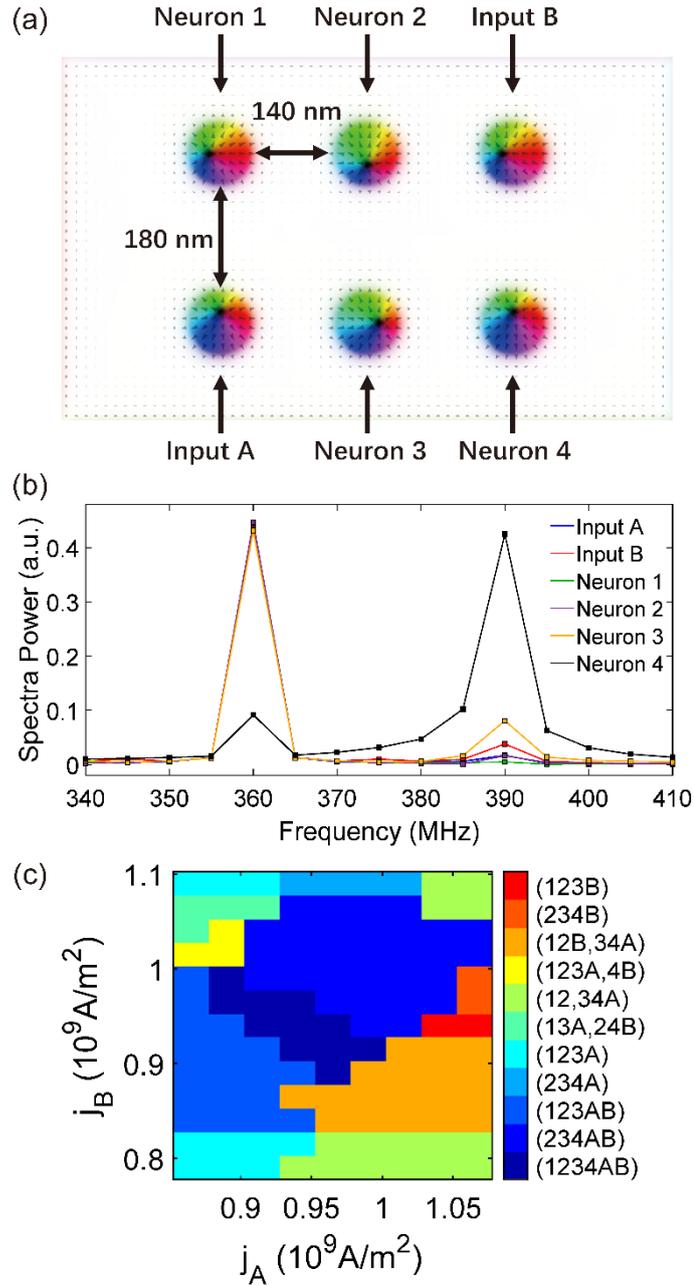

Fig.5 (a) Configuration of the neural network with two input oscillators and four neuron oscillators. ($j_1$=0.865× $10^9$A/m$^2$, $j_2$=0.925× $10^9$A/m$^2$, $j_3$=0.955× $10^9$A/m$^2$, $j_4$=1.02× $10^9$A/m$^2$) (b) The output frequencies of the oscillators. All the oscillators except Neuron 4 are synchronized. ($j_A$=0.94× $10^9$A/m$^2$, $j_B$=0.89× $10^9$A/m$^2$) (c) Synchronization map when both $j_A$ and $j_B$ are swept. A total of 11 different locked outputs can be achieved.